\documentclass[final,1p,times]{elsarticle}

\usepackage{amssymb} 
\usepackage{amsmath} 
\usepackage{wasysym}
\usepackage{float}
\usepackage{import}
\usepackage{siunitx}
\usepackage[hidelinks]{hyperref}
\usepackage{hyperref}
\usepackage{tabularray}
\usepackage{placeins}
\usepackage{enumitem}

\DeclareSIUnit{\pixel}{pixel}
\DeclareSIUnit{\sample}{samples}
\DeclareSIUnit{\frame}{frames}

\hypersetup{
  pdftitle={The tube transducer as a novel source for power ultrasound: A case study in delamination of graphite coating from lithium-ion battery anode},
  pdfauthor={Shida Li, Paul Daly, Ben Jacobson, Joshua Cooke, Chunhong Lei, Andrew P. Abbott, Andrew Feeney, Paul Prentice}}

\begin{document}
\begin{frontmatter}
\title{The tube transducer as a novel source for power ultrasound: A case study in delamination of graphite coating from lithium-ion battery anode}
\author[inst1]{Shida Li\corref{equalcontrib}} 
\author[inst1]{Paul Daly\corref{equalcontrib}}
\author[inst1]{Ben Jacobson}
\author[inst1]{Joshua Cooke} 
\author[inst2]{Chunhong Lei}
\author[inst2]{Andrew P. Abbott}
\author[inst1]{Andrew Feeney}
\author[inst1]{Paul Prentice\corref{cor1}}
\cortext[equalcontrib]{These two authors contributed equally to this work.}
\cortext[cor1]{Corresponding author}
\ead{paul.prentice@glasgow.ac.uk}
\address[inst1]{James Watt School of Engineering, University of Glasgow, Glasgow G12 8QQ, UK}
\address[inst2]{School of Chemistry, University of Leicester, Leicester LE1 7RH, UK}

\begin{abstract}
Developing high throughput applications of sonochemistry and sonoprocessing is an outstanding ultrasonic engineering challenge that continues to limit widespread industrial adoption. Conventional mass-produced Langevin-based technologies, such as the sonotrode or cleaning bath transducers, are not particularly well suited to treating large liquid volumes or flow-based systems, with a compromise between cavitation intensity and distribution through liquid bulk typically required. We report on the development of a tube transducer from a single element radially poled tubular piezoceramic, excited to generate an axially focused field. High-speed imaging and sonochemiluminescence are used to characterise the cavitation generated, which is also compared to the well-known activity at the tip of a sonotrode. Tube transducer and sonotrode sonications are then assessed for the material recycling application of graphite coating delamination from lithium-ion battery anode, both for intact and flaked anode sheets. The findings show that the tube transducer generates cavitation at sonotrode-like intensities or higher but distributed throughout the bore of the tube, with peak activity at the central axis. Prospects for developing tube transducer technology for high throughput flow-based applications are discussed. 
\end{abstract}

\begin{highlights}
\item An axially focused tube transducer is developed for power ultrasound applications
\item The cavitation generated is characterised and compared to that of a sonotrode
\item Tube cavitation is more intense for the same input power, and distributed through the bore
\item Lithium-ion battery anode delamination is studied for various sonotrode and tube configurations
\item Prospects for high throughput, flow-based sonoprocessing are discussed
\end{highlights}

\end{frontmatter}

\section{Introduction} \label{sec:intro}
The Langevin transducer is the dominant source device in power ultrasonics \cite{perez2020numerical, zhang2022investigation}, including for sonochemistry \cite{thompsonSonochemistryScienceEngineering1999, seidiAnalyticalSonochemistryDevelopments2012} and sonoprocessing \cite{morton2021new, Mer22,priyadarshiRoleShockWaves2025} applications. The initial design by Langevin and Chilowsky employed quartz as the piezoelectric actuator \cite{duck2022paul}. Modern Langevins typically consist of a stack of piezoceramic rings, sandwiched between front and back masses, to promote transmission through the transducer front-face. A central bolt pre-stresses the ceramics to reduce fracturing during expansion under tension \cite{Zha16, LiX21, LuX17}. 

The Langevin transducer is most commonly (and commercially) deployed in one of two configurations: the sonotrode \cite{jacobson2023mechanistic, lei2021lithium, lei2024effect} or the ultrasonic cleaning bath \cite{he2015recovery, marshall2020disassembly}. For the former, the vibrations generated by the transducer are transmitted through a detachable metal probe---the front mass, which is often stepped or tapered for amplification \cite{peshkovsky2007matching, khairiyah2023design}---to the tip of the probe, immersed within the liquid medium \cite{yusuf2021characterising}. Tip vibrations, typically in the range of several to hundreds of microns, generate intense cavitation activity in the liquid directly below the tip, although streaming effects often circulate smaller cavitating bubble clusters into the liquid bulk \cite{yusuf2021characterising, morton2023dual}. For the latter cleaning bath configuration, a number of Langevin transducers are bonded to the outer surfaces of a generally cuboidal metal vessel, which typically has a capacity of several litres \cite{Mer22, adamou2024ultrasonic}. The cavitation activity generated is distributed through the bulk of the liquid, but at a much reduced intensity compared to that at the vibrating tip of a sonotrode \cite{he2015recovery, xiao2021ultrasound, lei2021lithium}.

Many other sonoreactor designs are available or have been reported, that seek to address the inherent intensity versus distribution limitations of the two basic configurations, described above. The ultrasonic cup \cite{teixeira2014fast, santos2018ultrasound}, for example, is an attachment option for an inverted sonotrode device, that effectively couples a large tip diameter directly to the liquid contained within the cup. This achieves superior cavitation distribution through a larger volume, than the localised cavitation generated with a narrower tip attachment dipped into a vessel of liquid, but inevitably at lower intensity due to the larger emitting surface. Another approach is to mount multiple bath-type transducers to cylindrical or polygonal shaped vessels, to achieve a focusing effect for intensification of the cavitation activity \cite{Gog03, Lip19, hodnettChapter8Measurement2023}. Incorporating many bath-type transducers into a reactor design also allows for a larger volume of liquid to be addressed, and offers options such as multiple frequency sonications, but at the expense of cost and operational complexity.

In this paper, we describe the development of a transducer based on a radially poled tubular piezoceramic element, with the intention of addressing some of the limitations of conventional Langevin-based reactors. High-speed imaging (HSI) and sonochemiluminescence (SCL) are used to characterise the cavitation activity in the liquid contained within the bore of the tube transducer. Moreover, the performance of a tube transducer for delaminating graphite coating from lithium-ion battery (LiB) anode is assessed relative to that of a conventional sonotrode device, in a comparable exposure geometry. This sonoprocessing application has been recently reported for both cleaning bath \cite{he2015recovery, xiao2021ultrasound} and sonotrode configurations \cite{lei2021lithium, lei2024effect}. Lei \textit{et al.} \cite{lei2021lithium, lei2024effect} describes the use of a $\qty{1250}{\watt}$, $\qty{20}{\kilo\hertz}$ sonotrode with a $\qty{20}{\milli\metre}$ diameter tip attachment for sonicating LiB anode (and cathode) sheets submerged in deionised water, to remove the bonded powdered active material from the copper (and aluminium) foil current collector. For a tip-to-anode distance of $\qty{5}{\milli\metre}$ and a $\qty{3}{\second}$ sonication at an intensity of $\qty{120}{\watt \per \centi \metre \squared}$, the authors reported the removal of electrode graphite coating from an approximately circular area with a diameter  ${\approx}\qty{10}{\milli \metre}$. The size and shape of this ``treatment zone'' is attributable to the conical bubble structure that forms at the tip of a sonotrode. Those results illustrate the requirement for a sonotrode tip to be in close proximity to the target material and the small area typically treated during sonication, which are two clear limitations in the development of upscaled processing. Indeed, Lei \textit{et al.} also describes a novel blade-type sonotrode attachment, for rapid delamination of complete LiB electrode sheets, but which still required positioning to within $\qty{5}{\milli\metre}$ of the sheet.

In this work, we subject intact anode sheet sections to sonications from a sonotrode with a $\qty{20}{\milli\metre}$ tip and the tube transducer. To further explore the capabilities of the tube transducer, we also treat anode flakes and assess graphite coating removal efficiency via ``percentage remaining mass'', post-sonication. Tube transducer performance is compared to various sonotrode configurations, where flakes were sonicated within a cylindrical vessel with dimensions that match those of the tube transducer, with the tip at various vertical positions. The relative strengths of each exposure configuration are described and prospects for incorporating tube transducers into flow configurations for high throughput sonoprocessing and sonochemistry applications, discussed.

\section{Materials \& Methods}
\subsection{Tube transducer}\label{sec:methods_tube_transducer}
The actuator of the tube transducer is a single element, radially poled, piezoceramic tube with outer and inner diameters of $\qty{63.4}{\milli\metre}$ and $\qty{55.6}{\milli\metre}$ and a height of $\qty{30.3}{\milli\metre}$ (F3260232, CTS Ferroperm, Denmark). The piezoceramic material is PZ26, supplied with a conductive silver electrode coating on the inner and outer curved walls. Wiring is soldered to the tube and polyurethane (Urethane Conformal Coating, CRC Industries, USA) and epoxy resin (AbleStik 45, Loctite, USA) coatings are applied as electrically insulating and cavitation resistant barriers. The tube transducer was mounted within a custom designed, 3D printed housing for the purpose of studying the cavitation generated in liquid contained within the bore of the tube. An exploded view drawing of the transducer housing is presented in figure~\ref{fig:platform}~(b) where the piezoceramic tube is black and the printed housing components, grey. Two acrylic discs serve as observation windows for optical imaging. Figure~\ref{fig:platform}~(c) shows the tube transducer within the assembled housing, made water-tight with silicone sealant (820, No Nonsense, Screwfix, UK).

\begin{figure*}[t] 
\centering 
\includegraphics[]{fig1_platform.png}
\caption{Schematics and photographs of the tube transducer and sonotrode exposure configurations, and the experimental arrangement used for the high-speed imaging. (a)~Equipment items include (1)~tube transducer and housing, (2)~ring illumination, (3)~laser illumination, (4)~high-speed camera, (5)~signal generator, (6)~power amplifier and (7)~electrical impedance matching transformer. (b)~An exploded view drawing  of the tube transducer within its housing. (c)~A tube transducer with an intact LiB anode section mounted horizontally in a custom-made bracket (see figure~\ref{fig:LiB}~(b)) and marked with a red area. (d)~An exploded view drawing of the cylindrical sonotrode vessel. (e)~A schematic of the sonotrode configured to treat an intact LiB anode section. (f)~A schematic of the sonotrode with the tip positioned centrally within a cylindrical vessel, used to treat flakes of LiB anode. Red, green and blue arrows denote the vertical position of the sonotrode tip in the top, centre and bottom positions.}
\label{fig:platform}
\end{figure*}

\begin{figure*}[t]
    \centering
    \includegraphics[]{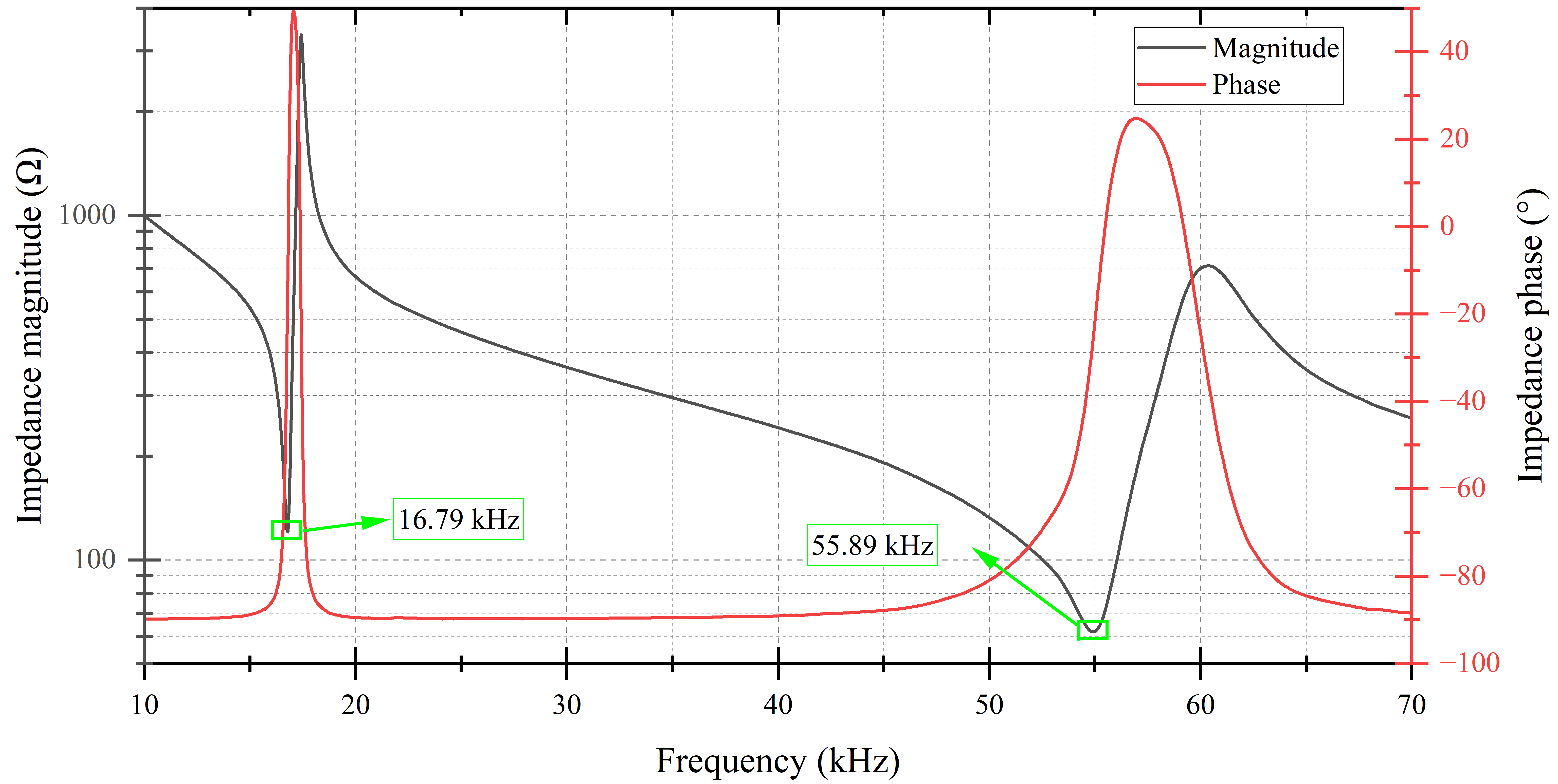}
    \caption{Measured electrical impedance of the tube transducer with frequencies of first radial and axial modes noted.}
    \label{fig:Ze_spectrum}
\end{figure*}

The impedance of the tube transducer was measured using an electrical impedance analyser (4294A, Keysight Technologies, CA, USA). The spectrum is shown in figure~\ref{fig:Ze_spectrum} where the resonance of the fundamental radial mode is apparent at $\qty{16.79}{\kilo\hertz}$. This mode was targeted for generating an axially focused field for high intensity cavitation. The tube transducer was therefore excited by a sinusoidal voltage signal at this frequency for all results presented. Amplification and impedance matching are described in section~\ref{sec:methods_tube_excitation}, below. A second resonance frequency appearing at $\qty{55.89}{\kilo\hertz}$ is the first axial mode, but the pressure field generated inside the tube bore at this frequency does not have the desired distribution and so was not applied in this study.

\subsection{Tube transducer excitation}\label{sec:methods_tube_excitation} 
A signal generator (DG4102, RIGOL Technologies, Beijing, China), shown in figure~\ref{fig:platform}~(a) equipment item~5, was used to generate a continuous $\qty{16.79}{\kilo\hertz}$ sinusoidal voltage signal of $\qty{2}{\second}$ duration, that was input to a ${+}\qty{55}{\decibel}$ power amplifier (1040L, Electronics and Innovation, USA) (figure~\ref{fig:platform}~(a) equipment item~6) to excite the tube transducer to generate a sonication. The $\qty{50}{\ohm}$ output of the amplifier was matched to the tube transducer by an electrical impedance matching transformer (B66397G0000X187, ETD59/31/22, N87, TDK EPCOS, Germany, equipment item 7) with a 1:5.7 turns ratio. A separate $\qty{1}{\volt}$ square wave signal triggered the high-speed camera (figure~\ref{fig:platform}~(a) equipment item~4). A range of tube transducer input powers, determined by the voltage amplitude of the $\qty{16.79}{\kilo\hertz}$ sinusoid set on the signal generator, were tested during preliminary experiments. Two powers were selected to provide an equitable comparison with the sonotrode (described below, section~\ref{sec:methods_sonotrode}) and to demonstrate the capabilities of the tube transducer. These are $\qty{55}{\watt}$, to match the sonotrode input power used exclusively throughout this work and $\qty{106}{\watt}$, the maximum tube transducer input power achievable by the equipment when the tube is filled with deionised water. The input power measurement method is detailed in section~\ref{sec:experimental_procedure}.

\subsection{Sonotrode}\label{sec:methods_sonotrode}
The cavitation generation and sonoprocessing performance of the tube transducer was compared to that of a commercial sonotrode (P100, Sonic Systems Ltd, UK), represented schematically in figure~\ref{fig:platform}~(e \& f). This sonotrode has an operating frequency of $\qty{20}{\kilo\hertz}$ and a tip diameter of $\qty{20}{\milli\metre}$ was used, matching that of Lei \textit{et al.} \cite{lei2021lithium, lei2024effect}. Sonotrode sonications were studied in two different reservoir types. The first, shown in figure~\ref{fig:platform}~(e), was used for the sonoprocessing of intact LiB anode sections. The second was a cylindrical vessel, with dimensions matching that of the tube transducer (see section~\ref{sec:methods_tube_transducer}) and shown in figure~\ref{fig:platform}~(d \& f), used for the sonication of anode flakes to provide an equitable comparison of the capabilities of each transducer type. An aperture in the wall of the vessel receives the tip of the sonotrode, with three vertical tip positions tested as represented by coloured arrows in figure~\ref{fig:platform}~(f): aligned to the top, centre and bottom of the vessel, approximately $\qty{5}{\milli\metre}$ above the lowest point.

For this system, the tip displacement amplitude is user selected and maintained during the sonication, irrespective of the acoustic load, while electrical power from the driver is automatically varied to meet the displacement amplitude demands. Transducer displacement from $\qty{0}{\micro \metre}$ to $\qty{16}{\micro \metre}$ can be selected, corresponding to transducer electrical input powers between $\qty{0}{\watt}$ to $\qty{120}{\watt}$, according to the manufacturer's specifications. For this work, the maximum electrical input power that was attained for all sonotrode sonications in deionised water, was measured to be $\qty{55}{\watt}$.

\subsection{Lithium-ion battery (LiB) anode samples} \label{sec:methods_LiB}
\begin{figure*}[t]
\centering
\includegraphics[]{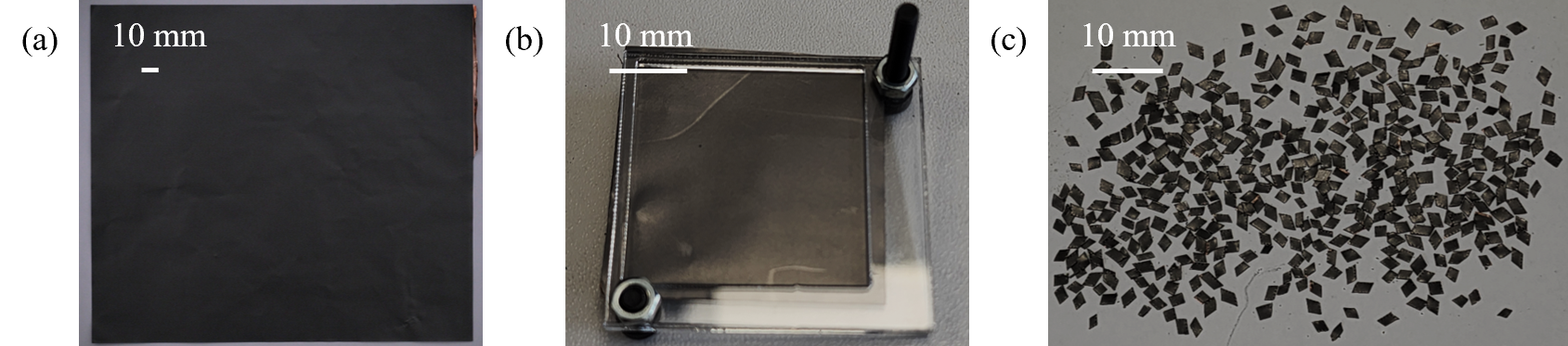}
\caption{(a)~A complete  LiB anode sheet. (b)~An intact LiB anode section mounted within a bracket. (c)~LiB anode flakes with an effective area of $\qty{17}{\centi \metre \squared}$.}
\label{fig:LiB}
\end{figure*}

Complete LiB anode sheets measuring approximately $\qtyproduct{23 x 20}{\centi\metre}$, composed of a $\qty{15}{\micro\metre}$ thick copper foil current collector coated on both sides with $\qty{70}{\micro\metre}$ layer of active material, figure~\ref{fig:LiB}~(a), were sourced from unused Nissan Leaf batteries. The active material coat is composed of graphite (with an average particle diameter of $\qty{15}{\micro\metre}$) and \textit{ca}. 4.5 wt\% CMC/SBR binder \cite{lei2021lithium}. The surface density of the LiB anode was measured as $\qty{0.030}{\gram\per\centi\metre\squared}$ using high precision scales (PCE-BSK~310, PCE Instruments) with an accuracy of $\qty{0.001}{\gram}$.

The experimental studies of Lei \textit{et al.} \cite{lei2024effect} show that a sonotrode with a $\qty{20}{\milli\metre}$ diameter tip at a distance of $\qty{5}{\milli \metre}$ from the surface of an intact LiB cathode section, removes graphite coating from a circular area with a diameter of approximately $\qty{10}{\milli\metre}$, from the side of the cathode facing the tip. For the current work, an intact section of LiB anode was held within a mounting bracket, figure~\ref{fig:LiB}~(b), and suspended horizontally beneath either the sonotrode tip, figure~\ref{fig:platform}~(e), or across the centre of the tube transducer, figure~\ref{fig:platform}~(c). Within the bracket the anode section has an exposed area of $\qtyproduct{3 x 3}{\centi\metre}$, with the section cut marginally larger for secure mounting. The sonotrode was operated at $\qty{55}{\watt}$ for $\qty{2}{\second}$. for tip-to-anode distances of $\qty{15}{\milli \metre}$ and $\qty{5}{\milli \metre}$. The effects of tube transducer sonication on the intact section, when operating at $\qty{55}{\watt}$ and $\qty{106}{\watt}$ input powers for $\qty{2}{\second}$, were also tested. Five intact anode sections were sonicated for each of these four transducer configurations (two sonotrode tip-to-anode distances and two tube transducer input powers), as presented in \textit{Results} section~\ref{sec:results_lib_section}.

In an attempt to reduce the dependence of sonoprocessing effects on the spatial distribution of the cavitation activity generated by the two transducer types used in this study, we also assessed the graphite coating delamination performance with LiB anode flakes, as seen in figure~\ref{fig:LiB}~(c). A number of approaches to material recovery from end-of-life LiB anodes include mechanical crushing or shredding, suggesting the use of flakes is both valid and of practical interest \cite{gaines2014future, kosenkoReviewExistingStrategies2023}. Batches of flakes of length and width less than $\qty{2}{\milli\metre}$ were cut from anode sheet, figure~\ref{fig:LiB}~(c). Flake batches of effective areas up to $\qty{24}{\centi \metre \squared}$ were tested, as detailed in \textit{Results} section~\ref{sec:results_effect_area}. Graphite coating removal was assessed by weighing dry anode flakes from each batch before and after each $\qty{2}{\second}$ sonication. Five sample batches were tested for each of the five transducer configurations investigated (the sonotrode with three different tip positions and the tube transducer at the two input powers), with data presented as the average $\pm$ standard deviation. Each batch of flakes was added to the water within either the bore of the tube transducer or the cylindrical vessel (prior to tip insertion), and allowed to settle to the bottom before the sonication was initiated, as can be seen in figures~\ref{fig:camera_image_4cm2} and \ref{fig:camera_image_17cm2}.

\subsection{Experimental procedure} \label{sec:experimental_procedure}
Deionised water at room temperature was the host liquid medium for all investigations (with the exception of the SCL study, detailed in section~\ref{subsec:methods_SCL}). Any observed differences in cavitation intensity or graphite coating delamination performance can therefore be attributed to the transducer configuration or operating conditions.

Electrical power input into the tube transducer and sonotrode were measured by using a digital multimeter (117 TRUE RMS, Fluke, USA). The setup involves connecting the digital multimeter between the tube transducer and the transformer, and between the sonotrode and the control panel.

Post-sonication imaging of intact anode sheet sections and batches of flake samples were taken with a surface microscope (SMZ-171, Motic, China; maximum magnification $\times 50$).

\subsection{Optical imaging}\label{sec:methods_imaging}
\subsubsection{High-speed imaging (HSI)}\label{subsec:methods_HSI}
HSI of the cavitation inside the tube transducer and the cylindrical sonotrode vessel during a sonication was captured with a FASTCAM SA-Z 2100~K (Photron, Bucks, UK) high-speed camera, figure~\ref{fig:platform}~(a) equipment item~4. The high-speed camera recorded the cavitation activity for the entire $\qty{2}{\second}$ duration of each sonication at $\qty[per-mode = symbol]{20000}{\frame \per \second}$, through a macro-lens (Milvus 100~mm f/2M, Zeiss, Oberkocken, Germany). At this frame rate, imaging was achieved over $\numproduct{1024 x 1024}$ pixels, with a resolution of $\qty{54}{\micro\metre\per\pixel}$. Dual illumination fixtures, shown in figure~\ref{fig:platform}~(a) equipment items~2 and 3, ensured full illumination of the tube transducer and cylindrical vessel interiors. The peripheral regions were illuminated by a $\qty{150}{\watt}$ halogen lamp (Thorlabs, Ely, UK) coupled with a fibre ring illuminator (FRI61F50, Thorlabs). Central regions were irradiated by synchronous $\qty{10}{\nano\second}$ laser pulses at $\qty{640}{\nano\metre}$ (CAVILUX Smart, Cavitar, Finland), coupled to a liquid light guide with a collimating lens attached to the output end.

\subsubsection{Sonochemiluminescence (SCL) observation}\label{subsec:methods_SCL}
SCL imaging was employed to qualitatively assess cavitation activity inside the tube transducer and the cylindrical sonotrode vessel. Sonications for each of the five transducer configurations (see sections~\ref{sec:methods_tube_excitation} and \ref{sec:methods_sonotrode}) were undertaken with the SCL working solution---$\qty{1}{\litre}$ deionised water with 0.1~M NaOH and $\qty{0.17}{\gram}$ of luminol \cite{lei2024effect}---in a dark room. Sonication causes luminescent phenol to react with hydroxyl radicals created by sonolysis to produce 3-aminophthalate, which emits blue light \cite{wang2022ultrasound}. It is noted that although SCL indicates activity associated with the generation of reactive species, this also indirectly indicates the intensity of local cavitation activity \cite{lei2024effect}. To generate sufficient luminescence, sonications of $\qty{10}{\second}$ durations were necessary. Images were captured using a Nikon~Z8 digital camera equipped with a NIKKOR~Z 24-120~mm f/4~s lens with a $\qty{10}{\second}$ exposure time; \textit{f}/4 aperture and 3200~ISO. No post-processing was applied to the captured SCL images.

Consequently, the original $\qty{24}{\bit}$ true colour RGB images were converted to $\qty{8}{\bit}$ greyscale images using the \texttt{rgb2gray} function in MATLAB \cite{themathworksinc.Rgb2grayConvertRGB}. There were no instances of 0~value greyscale levels throughout these converted images and so a filter was created in Photoshop (Adobe Inc., San Jose, CA, USA) that assigned a 0~value to pixels corresponding to the transducers, vessels and exterior areas (the non-liquid areas that do not emit SCL). Greyscale level histograms of these altered images were calculated using the \texttt{histogram} function in MATLAB (R2025a, MathWorks, MA, USA) and the values corresponding to the 0~level were nullified \cite{themathworksinc.HistogramHistogramPlot}. The histogram therefore represents the prevalence of $2^8 - 1 = 255$ possible greyscale levels within the cylindrical sonotrode vessel and the tube transducer, in the processed versions of figure~\ref{fig:SCL}~(a)--(e). The magnitude and prevalence of these greyscale levels correspond to the intensity and distribution of SCL, which in turn may be taken as an indicator of the intensity of the cavitation activity \cite{tiongComparisonSonochemiluminescenceImages2017}.

\subsubsection{Conventional imaging of graphite coating removal}\label{subsec:methods_camera_imaging}
Conventional video imaging at $\qty[per-mode = symbol]{120}{\frame \per \second}$ captured the graphite coating delamination from LiB flakes within the tube transducer and the cylindrical sonotrode vessel, during sonications. Again, a Nikon~Z8 digital camera equipped with a NIKKOR~Z 24-120~mm f/4~s lens was used. Imaging was undertaken over $\numproduct{3840 x 2160}$ pixels with 1/125~$\unit{\second}$ exposure time, \textit{f}/4 aperture and 3200~ISO. Example video data from both HSI and conventional imaging are available as \textit{Supplementary Materials}.

\section{Results}
The following results sections are organised to compare the cavitation generated by the tube transducer and the sonotrode. Their effectiveness in delaminating graphite coating from LiB anode, as intact sections and in flake form, is evaluated for the various sonication configurations described above.
\begin{itemize}
    \item Section~\ref{sec:results_hsi} presents HSI results of cavitation activity generated by the sonotrode at $\qty{55}{\watt}$, with the tip at three vertical positions within a cylindrical vessel, and by the tube transducer at two input powers ($\qty{55}{\watt}$ and $\qty{106}{\watt}$). This allows a comparison of the bubble distribution and structures that develop, during sonications with each configuration.
    \item Section~\ref{sec:results_scl} evaluates the spatial distribution and intensity of the cavitation activity via SCL imaging through qualitative observations and quantitatively, with greyscale histogram analysis. 
    \item Section~\ref{sec:results_lib_section} presents the graphite coating delamination observations for the sonotrode and the tube transducer on intact LiB anode sections. These sonotrode sonications were undertaken in the tank of figure~\ref{fig:platform}~(e), not the cylindrical vessel, such that the mounted anode section occupied a plane below the sonotrode tip, for comparison to the results of Lei \textit{et al.} \cite{lei2024effect}. Results for two tip-to-anode distances ($\qty{5}{\milli \metre}$ and $\qty{15}{\milli \metre}$) and the two input powers to the tube transducer are described.
    \item Section~\ref{sec:results_lib_flake} presents the observations of graphite coating removal from LiB anode flakes, with an effective area of $\qty{4}{\centi \metre \squared}$, for each of the five transducer configurations investigated by the optical imaging techniques, sections~\ref{sec:results_hsi} and \ref{sec:results_scl}. The results represent the conventional imaging taken during sonications and a comparison of post-sonication ``remaining mass percentage''. 
    \item Section~\ref{sec:results_effect_area} explores the effect of increasing anode flake effective area loading on the graphite coating delamination capability of the tube transducer, at $\qty{106}{\watt}$ input power. A threshold is identified, beyond which delamination efficiency reduces due to overloading. Conventional imaging of graphite coating delamination and percentage remaining mass data, for this threshold effective area load are finally presented in section~\ref{sec:results_lib_17cm2}.
\end{itemize}

\subsection{High-speed imaging (HSI) of the cavitation generated by sonotrode and tube transducers} \label{sec:results_hsi}
\begin{figure*}[t]
\centering
\includegraphics[]{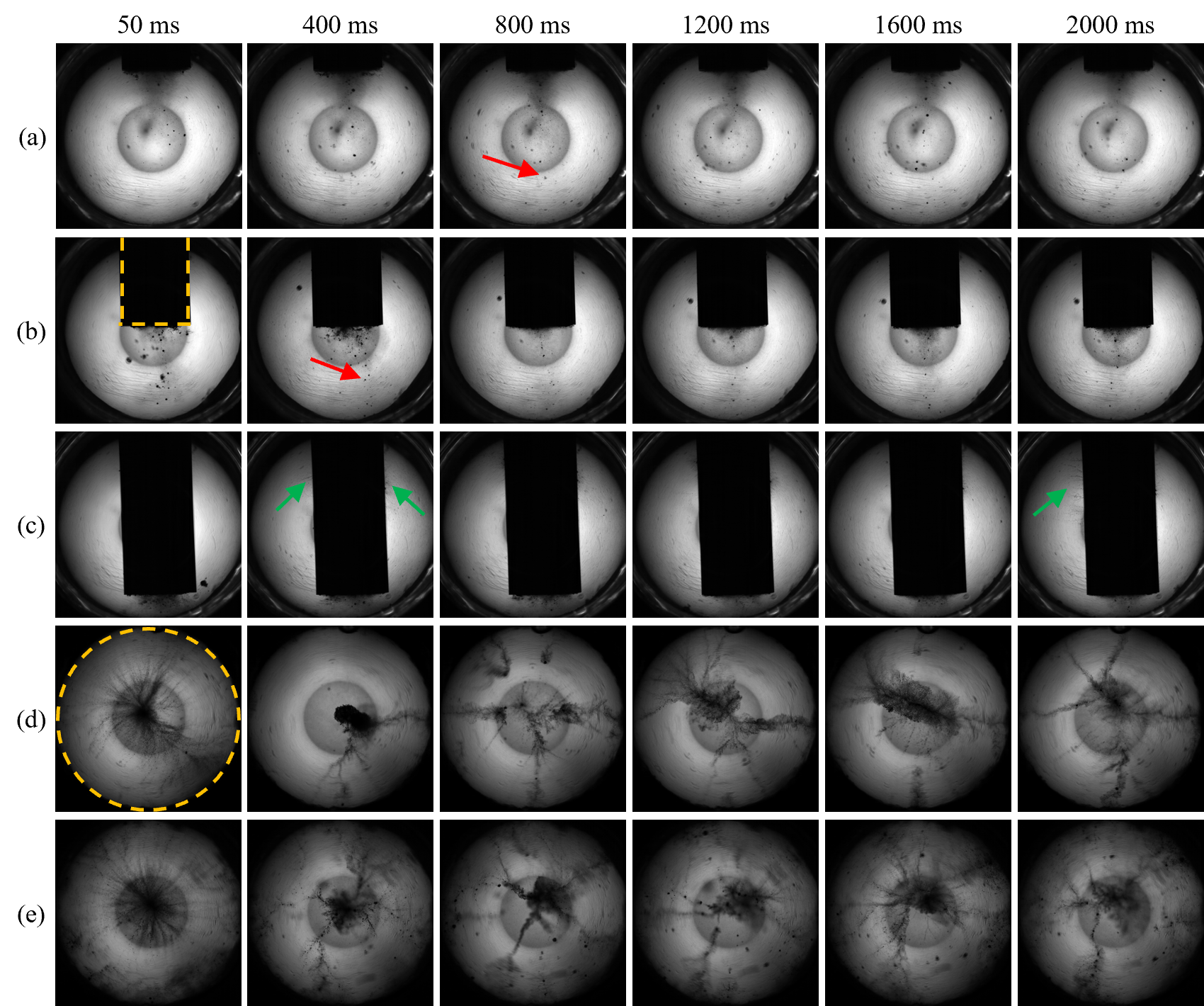}
\caption{Selected images from high-speed sequences captured at $\qty[per-mode = symbol]{20 000}{\frame \per \second}$, showing the acoustic cavitation generated by the sonotrode operating at $\qty{55}{\watt}$, with the tip in the (a)~top, (b)~central and (c)~bottom positions, and within the tube transducer at (d)~$\qty{55}{\watt}$ input power and (e)~$\qty{106}{\watt}$. Scale is provided by the $\qty{20}{\milli\metre}$ diameter sonotrode tip. All sonications were $\qty{2}{\second}$ in duration. The outlines of the sonotrode tip and the tube bore are shown by a dashed orange line in (b) and (d). Coloured arrows indicate small bubble clusters detached from the main conical structure (red) and cavitation along the shaft (green) for the sonotrode sonications.}
\label{fig:HSI}
\end{figure*}

Selected frames from HSI sequences captured at $\qty[per-mode = symbol]{20 000}{\frame\per\second}$ (corresponding to the $\qty{20}{\kilo\hertz}$ sonotrode frequency) of the cavitation generated by the sonotrode tip at the top, centre and bottom positions within the cylindrical vessel (as described in section~\ref{sec:methods_sonotrode}) are presented in figure~\ref{fig:HSI}~(a) to (c), respectively. The central circular shadow feature throughout figure~\ref{fig:HSI} is an artefact of the combined illumination fixtures (figure~\ref{fig:platform}~(a), equipment items~2 and 3). Results from HSI of the tube transducer are shown in figure~\ref{fig:HSI}~(d) and (e) at $\qty{55}{\watt}$ and $\qty{106}{\watt}$, respectively. These images are representative of the bubble structures that develop during the $\qty{2}{\second}$ sonications in each of the respective configurations. The outlines of the sonotrode tip and the tube bore are indicated by a dashed orange line in (b) and (d). 

Figure~\ref{fig:HSI}~(a) and (b) indicates that in the top and middle tip positions the classic cone shaped bubble structure forms, remaining in contact with the tip for the duration of the sonication, as has been extensively reported previously \cite{moussatov2003cone, bai2014generation, ma2017experimental}. Small oscillating clouds of bubbles (indicated by red arrows) occasionally detach from the main conical structure and migrate downwards through the cylindrical vessel under the action of acoustic streaming. In figure~\ref{fig:HSI}~(c), where the tip is closest to the bottom of the cylindrical vessel, cavitation activity is evident but the conical structure is not clearly formed in the constrained space. Cavitation bubbles are also observed along the shaft of the sonotrode probe (indicated by green arrows) as also previously reported for a tip sufficiently immersed within a liquid \cite{son2020geometric, garcia2023extensive, bampouli2024importance}.

Figure~\ref{fig:HSI}~(d) shows typical cavitation activity within the bore of the tube transducer at an input power of $\qty{55}{\watt}$. An experimental and numerical investigation of similar bubble structures that form under a ramped excitation voltage in an equivalent tube transducer has recently been reported \cite{metzgerRevisitingSubharmonicRoute2025}. Initially, within approximately $\qty{50}{\milli\second}$ of the sonication initiation, cavitation bubbles have rapidly formed many fine, radial filamentary structures throughout the bore of the tube, generating a densely packed cluster at the central, axial position. Bubbles and bubble clusters initially generated in the peripheral regions tend to migrate inwards, along these filamentary structures, ``feeding'' the central cluster. This activity is best perceived in movie format for the entire HSI sequence, available as \textit{Supplementary Materials}. From approximately $\qty{100}{\milli\second}$ of the sonication, the finer radial structures have combined to form more established filaments that, together with the central cluster, constitute the main activity for the remainder of the sonication.
 
In figure~\ref{fig:HSI}~(e), at the higher input power of $\qty{106}{\watt}$, cavitation bubbles fill the bore continuously throughout the sonication period. The bubbles exhibit a distinct spatial distribution and a larger central cluster forms. More filamentary structures also form between the outer regions and the central cluster than at the lower power. It can be seen in the movie format of this data, available as \textit{Supplementary Materials}, that these filament-like structures act as bubble migration channels and correspond to regions of intense cavitation activity, as confirmed by SCL images below, in section~\ref{sec:results_scl}.

These HSI observations demonstrate that the structure and distribution of cavitation bubbles generated by the sonotrode and tube transducers are notably different. For the sonotrode system, cavitation is localised to the tip according to the vertical position within the cylindrical vessel. In contrast, the filamentary structures generated by the tube transducer feed a cluster located on the central axis, channelling cavitation activity away from the inner surface of the transducer.

\subsection{Sonochemiluminescence (SCL) imaging of the sonotrode and tube transducers} \label{sec:results_scl}
\begin{figure*}[tp]
\centering
\includegraphics[]{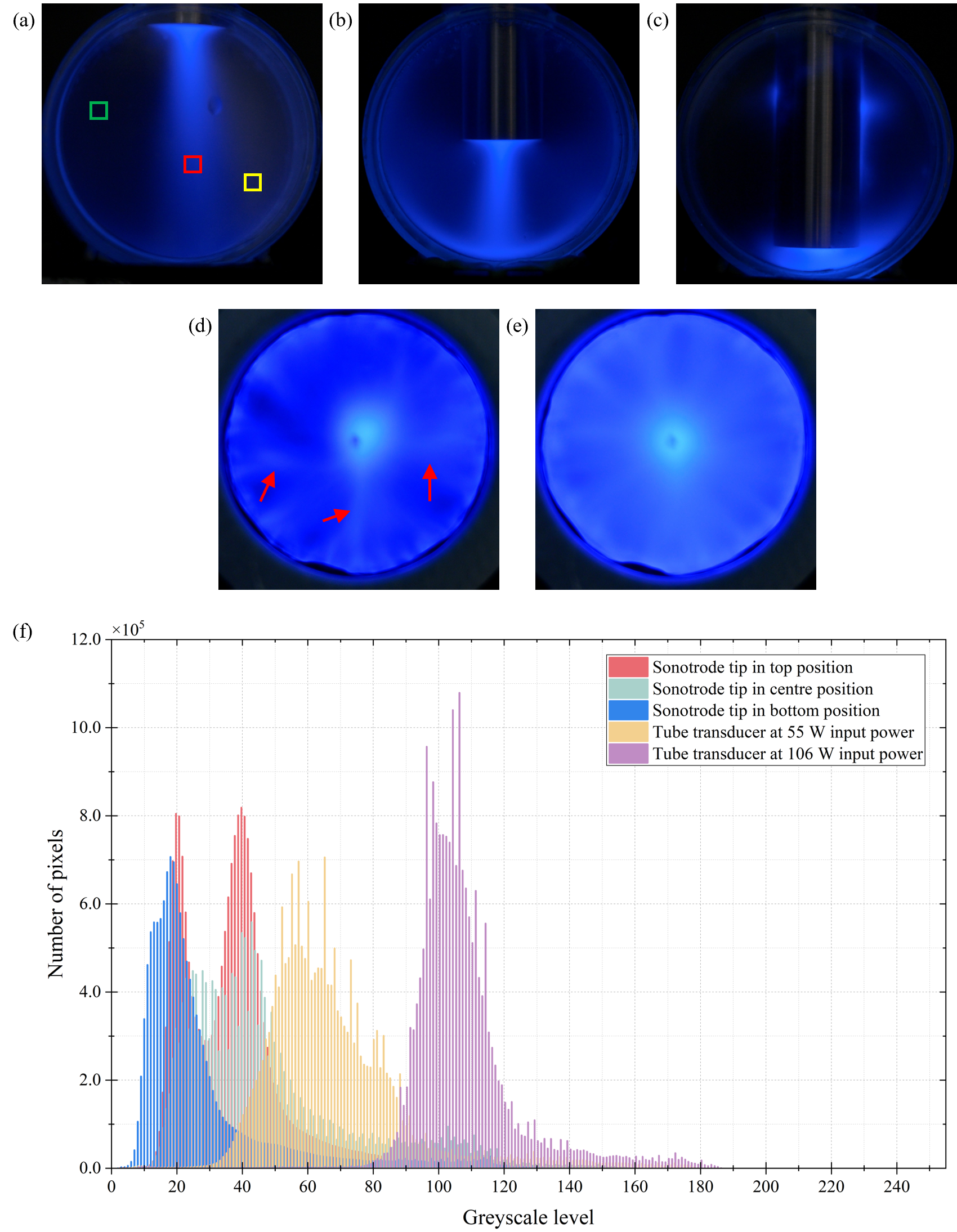}
\caption{SCL images obtained during $\qty{10}{\second}$ sonications for the sonotrode at $\qty{55}{\watt}$ with the tip in the (a)~top, (b)~centre and (c)~bottom positions within the cylindrical vessels, and the tube transducer at (d)~$\qty{55}{\watt}$ and (e) $\qty{106}{\watt}$ input power. (f)~Greyscale level histograms of (a)--(e) obtained by MATLAB greyscale processing. The boxes on (a) represent regions of greyscale values classified as invisible ($<30$, green), low intensity ($>30$, $<40$, yellow) and high intensity ($>40$, red). Red arrows in (d) indicate radial lines of luminescence, discussed in section~\ref{sec:results_scl}.}
\label{fig:SCL}
\end{figure*}

The SCL imaging of the sonotrode tip in its three positions and the tube transducer at each input power in figure~\ref{fig:SCL}~(a--e), align broadly to the cavitation activity seen during the HSI, despite being captured over longer $\qty{10}{\second}$ sonications. Figure~\ref{fig:SCL}~(a \& b) are typical of previous SCL-based studies of cavitation at a sonotrode tip, confirming that the most intense and sonochemically active cavitation is contained within the conical bubble structure, in contact with the sonotrode tip. The luminescence beyond this region, extending down through the cylindrical vessel, is attributable to the smaller detached bubble clusters also seen during the HSI (indicted by red arrows in figure~\ref{fig:HSI}~(a \& b)). During the sonication with the tip in the centre position, figure~\ref{fig:SCL}~(b), this extended luminescence region follows the lower inner surface of the cylindrical vessel. This is again consistent with the counter-rotating vortices known to be generated by the vibrating tip of a sonotrode, which will cause detached clusters to migrate along these trajectories, in this vessel geometry \cite{dahlemRadiallyVibratingHorn1999, priyadarshi2021situ}. The SCL imaging with the tip in its bottom position, figure~\ref{fig:SCL}~(c), indicates intense cavitation activity within the constrained liquid region, as well as along the shaft of the submerged probe.

An SCL image of the tube transducer with an input power of $\qty{55}{\watt}$ is shown in figure~\ref{fig:SCL}~(d). Notably, luminescence is generated from all regions within the tube bore and is brightest at the central axis, corresponding to the large cluster that was observed during HSI. The distribution is generally uniform away from this central region, although three radial lines of slightly higher intensity are present (indicated by red arrows), attributable to the filament structures observed by HSI, figure~\ref{fig:HSI}~(d). An increase in tube transducer input power to $\qty{106}{\watt}$ enhanced the overall blue light intensity, as seen in figure~\ref{fig:SCL}~(e), but the off-centre distribution is less homogenous. The central on-axis region is again brightest but high luminescence is also evident around the inner circumference of the tube and along radial regions, in accordance with the HSI of the cavitation activity, figure~\ref{fig:HSI}~(e).

The SCL distributions in figure~\ref{fig:SCL}~(a--e) are compared quantitatively in the image histogram of greyscale levels, shown in figure~\ref{fig:SCL}~(f). The magnitude and prevalence of these greyscale levels correspond to the intensity and distribution of SCL, which in turn may be considered as an indicator of cavitation activity intensity. Terms for greyscale level ranges were defined to aid analysis as ``invisible'' for levels below 30, ``low intensity'' for levels between 30 and 40, and ``high intensity'' for levels greater than 40. An example of each is marked up in figure~\ref{fig:SCL}~(a) as green, yellow and red boxes, respectively. The intensity and distribution of SCL is greatest for the sonotrode, when its tip is in the centre position where $\qty{21.06}{\percent}$ and $\qty{52.79}{\percent}$ of greyscale values are low intensity and high intensity, respectively.

The tube transducer generates brighter and more distributed luminescence at $\qty{55}{\watt}$ and $\qty{106}{\watt}$ input powers. For both input powers low and high intensity greyscale levels account for more than $\qty{99.60}{\percent}$ of SCL image pixels and the majority of these---$\qty{97.07}{\percent}$ for the $\qty{55}{\watt}$ configuration---are high intensity. While both input powers create unimodal distributions in the image histogram, at $\qty{106}{\watt}$ the distribution is narrower and its peak is positioned at a higher greyscale value. This indicates that at the higher input power, the sum of areas corresponding to intense cavitation activity increases and their distribution throughout the tube bore is more uniform. This quantitive comparison corroborates the visual comparison of figure~\ref{fig:SCL}~(d \& e) just described. The maximum greyscale level for the tube transducer, for both input powers, was 190. However, significantly more maximum level pixels were present at $\qty{106}{\watt}$ input power than at $\qty{55}{\watt}$, again indicating that the spatial extent of the intensely cavitating region is more significant at the higher power. The maximum sonotrode greyscale level was 120.

In summary, these results indicate that for equivalent input powers the cavitation activity generated by the tube transducer is more intense than that produced by the sonotrode and that the activity is more distributed through the equivalent volume.

\subsection{Graphite coating removal from intact lithium-ion battery (LiB) anode sections} \label{sec:results_lib_section}
\begin{figure*}[tp]
\centering
\includegraphics[]{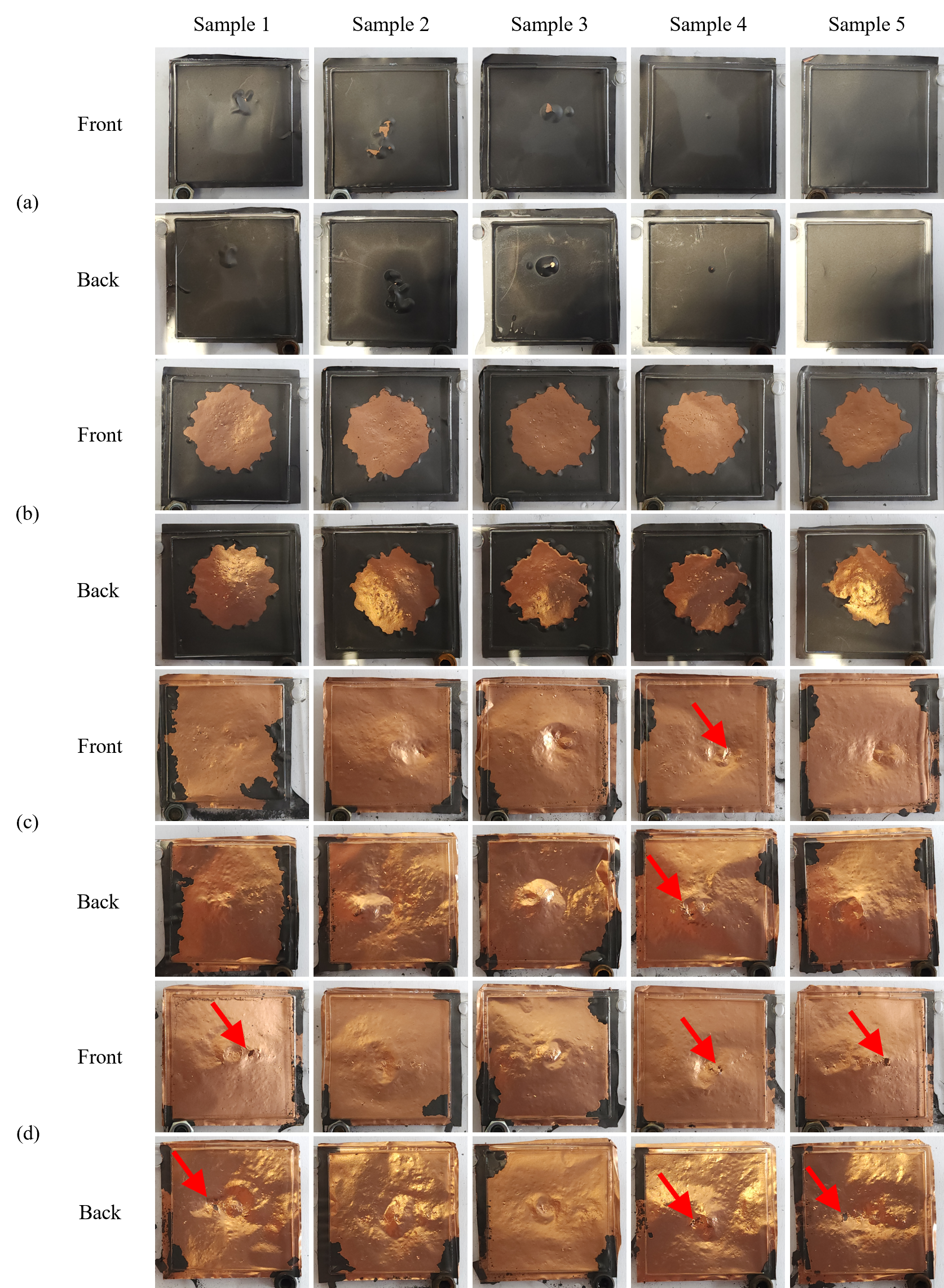}
\caption{The front and back surfaces of LiB anode sections mounted in brackets, with an exposed area of $\qtyproduct{3 x 3}{\centi\metre}$, following a $\qty{2}{\second}$ sonication for: Sonotrode at $\qty{55}{\watt}$ input power, with a separation distance of (a)~$\qty{15}{\milli \metre}$ and (b)~$\qty{5}{\milli \metre}$ between the tip and the anode section. Tube transducer at (c)~$\qty{55}{\watt}$ and (d)~$\qty{106}{\watt}$ input power. Each row shows five replicates treated with the same sonication conditions. Red arrows indicate puncture damage inflicted to the copper foil.}
\label{fig:sheet_sample}
\end{figure*}

\begin{figure*}[t]
\centering
\includegraphics[]{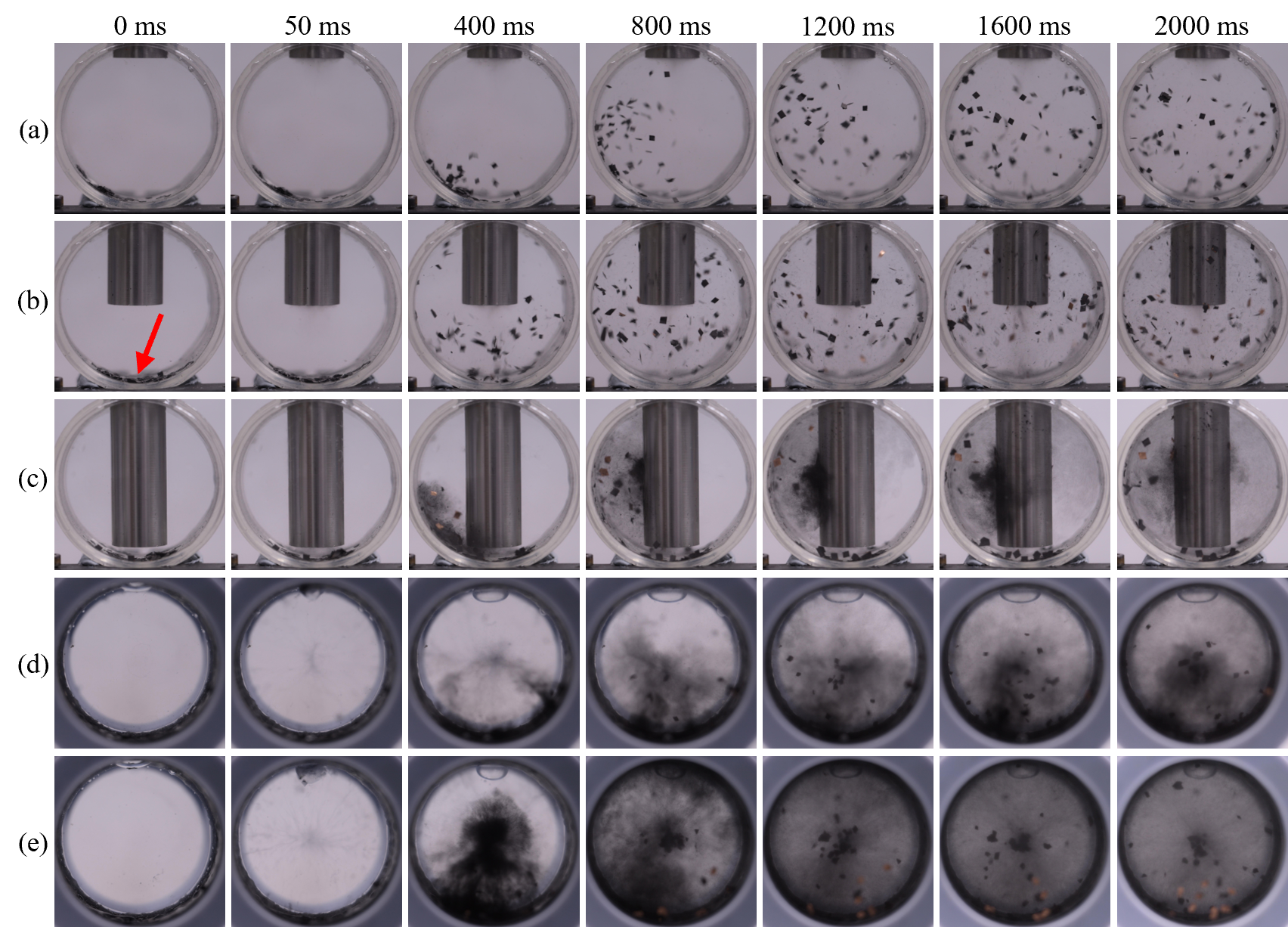}
\caption{Images from a conventional recording of graphite coating delamination from $\qty{4}{\centi\metre\squared}$ of LiB anode flakes, during a $\qty{2}{\second}$ sonication with the sonotrode tip in the (a)~top, (b)~centre and (c)~bottom position, and the tube transducer at (d)~$\qty{55}{\watt}$ and (e)~$\qty{106}{\watt}$ input power. Scale is provided by the $\qty{20}{\milli\metre}$ diameter sonotrode tip. The red arrow in (b) indicates the anode flakes settled at the bottom of the cylindrical vessel prior to sonication.}
\label{fig:camera_image_4cm2}
\end{figure*}

The surfaces of intact LiB anode sections, with a nominal area of $\qtyproduct{3 x 3}{\centi\metre}$, following a $\qty{2}{\second}$ sonication are shown in figure~\ref{fig:sheet_sample}, for the four transducer configurations described in section~\ref{sec:methods_LiB}. Figure~\ref{fig:sheet_sample}~(a \& b) show sections following treatment by the sonotrode transducer for tip-to-anode distances of $\qty{15}{\milli \metre}$ and $\qty{5}{\milli \metre}$, respectively. The effects of tube transducer sonication on sections, when operating at $\qty{55}{\watt}$ and $\qty{106}{\watt}$ input powers, are presented in figure~\ref{fig:sheet_sample}~(c \& d). Each row in figure~\ref{fig:sheet_sample} shows five replicates treated under the same transducer conditions and the labels ``Front'' and ``Back'' indicate upwards and downwards facing sides of the section as per figure~\ref{fig:platform}~(c \& e), with respect to the sonotrode tip.

In samples~1 to 4 of figure~\ref{fig:sheet_sample}~(a) the graphite coating has bulged away from the copper current collector on both the front and back sides, while on the front side of samples~2 and 3 these features have ruptured. These observations indicate that the adhesive bond between the active layer of graphite coating and the copper current collector has been broken. In all instances of figure~\ref{fig:sheet_sample}~(b), for a tip-to-anode distance of $\qty{5}{\milli \metre}$, an approximately circular area of graphite coating has been removed from both sides of the section, to expose the copper foil. Image analysis in MATLAB determined that the diameter and area of the delaminated zone are approximately $\qty{2.2}{\centi \metre}$ and $\qty{3.8}{\centi \metre \squared}$. 

The intact section delamination capabilities of the tube transducer, at the two input powers, are shown in figure~\ref{fig:sheet_sample}~(c \& d). Graphite coating has been removed from nearly all of the $\qtyproduct{3 x 3}{\centi\metre}$ areas and, in most cases, even the anode surface covered by the mounting bracket. Lei \textit{et al.} suggests that intense bubble collapse shockwaves enable delamination by both peeling off of the graphite coating layer and flexing the electrode material to induce stress fracturing, while cavitation bubble jetting creates pits in the foil \cite{lei2021lithium, lei2024effect}. The copper foil in figure~\ref{fig:sheet_sample}~(c \& d) has also visibly buckled owing to the intensity of cavitation activity, and jet pitting has caused punctures, which are most prominent in samples~1 and 5 of figure~\ref{fig:sheet_sample}~(d), arrowed red.

These observations show graphite coating removal by the sonotrode is reduced when the distance between the sonotrode tip and anode section is increased. Even when the tip-to-anode separation distance is small however, the treatment zone is limited to $\qty{3.8}{\centi \metre \squared}$ (the tip area, approximately). In contrast, the tube transducer was not limited by the positioning of the sample and removed graphite coating from effectively all of the $\qtyproduct{3 x 3}{\centi\metre}$ surface of the section.

\subsection{Graphite coating removal from anode flakes of \texorpdfstring{$\qty{4}{\centi\metre\squared}$}{4-cm-squared} effective area} \label{sec:results_lib_flake}
\begin{figure*}[t]
\centering
\includegraphics[]{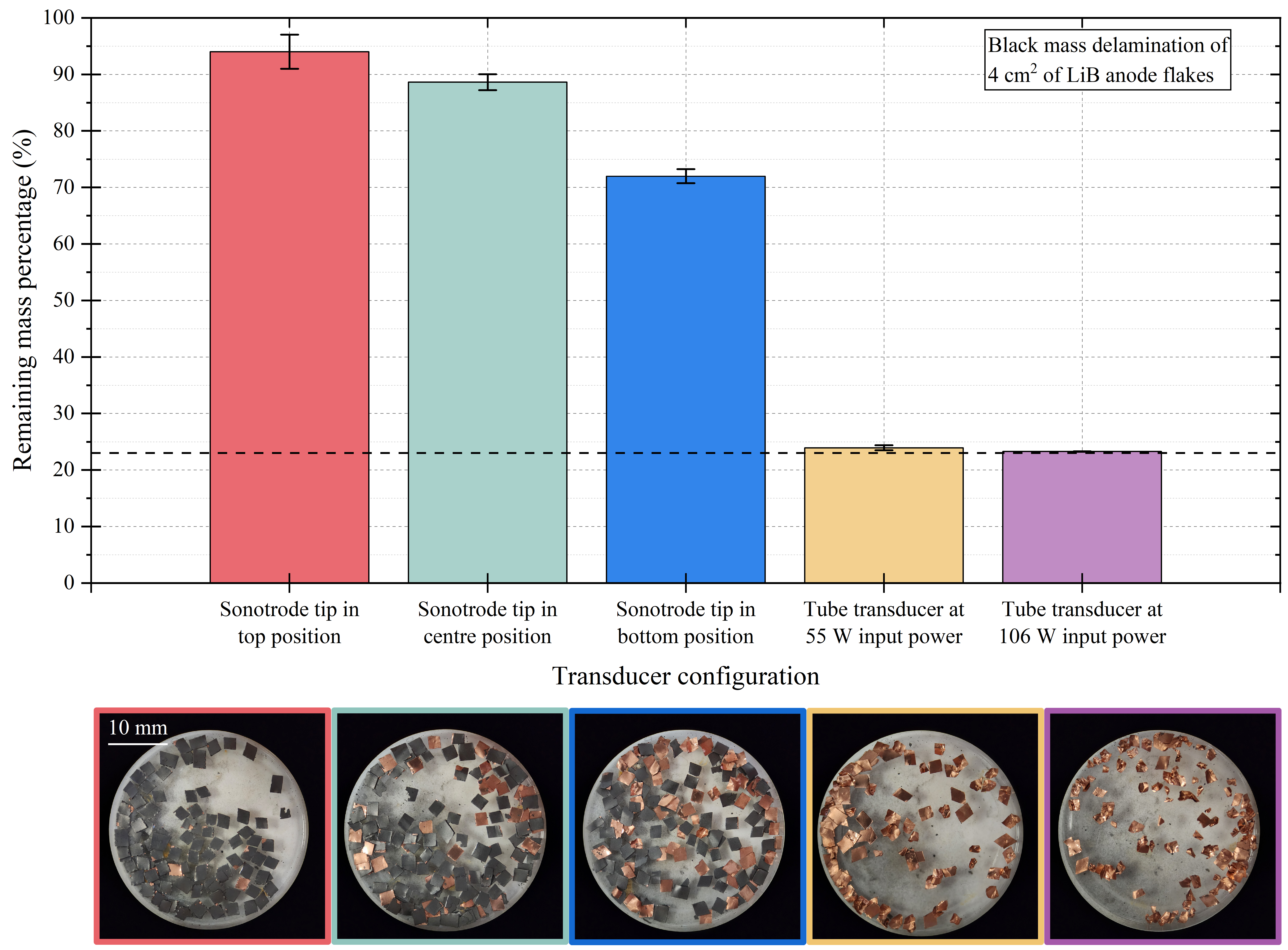}
\caption{Remaining mass percentage of LiB anode flakes with a $\qty{4}{\centi\metre\squared}$ effective area following $\qty{2}{\second}$ sonications in each of the five configurations. Complete graphite coating removal is indicated by a dashed line at $\qty{23}{\percent}$. Also shown are pictures of the flake batches, post-sonication, for each configuration.}
\label{fig:lib_compare_sonotrode_and_tube_4cm2}
\end{figure*}

To further test and (to some degree) mitigate the significant differences between the cavitation distributions generated by the sonotrode and tube transducers, LiB anode flakes were cut from $\qty{4}{\centi\metre\squared}$ of anode sheet. This allows for comparison with the sonotrode treatment described above, where $\qty{3.8}{\centi\metre\squared}$ of graphite coating was liberated from each side of a section (with the area rounded up for consistency with the $\qty{1}{\centi\metre\squared}$ increments of the flake loading investigation, section~\ref{sec:results_effect_area}, below). Graphite coating delamination from batches of LiB anode flakes with an effective area of $\qty{4}{\centi\metre\squared}$ is represented in figure~\ref{fig:camera_image_4cm2} for each of the five transducer configurations investigated, over $\qty{2}{\second}$ sonications. Figure~\ref{fig:camera_image_4cm2}~(a--c) depicts anode flake treatment by the sonotrode, at $\qty{55}{\watt}$ input power, with its tip in the top, centre and bottom positions, while figure~\ref{fig:camera_image_4cm2}~(d \& e) show the process in the tube transducer at $\qty{55}{\watt}$ and $\qty{106}{\watt}$ input powers.

Prior to sonication by the sonotrode, the flakes have settled to the bottom of the cylindrical vessel, as indicated by a red arrow in figure~\ref{fig:camera_image_4cm2}~(b). Figure~\ref{fig:camera_image_4cm2}~(a \& b) show that sonotrode sonications with the tip in the top and centre positions lift and disperse the flakes throughout the liquid volume. The degree of graphite coating delamination however, apparent by darkening of the water as graphite powder is liberated, is slight. Delamination is more evident in figure~\ref{fig:camera_image_4cm2}~(c), with the sonotrode tip in the bottom position, close to the flakes. Graphite coating is liberated from the flakes mainly in the constrained volume below the sonotrode tip. A few flakes are displaced by the acoustic streaming and may be subject to delamination by the cavitation along the shaft, as identified by the HSI and SCL observations, later in the sonication. Acoustic streaming also propels a proportion of the liberated powder upwards through the vessel, where it also collects preferentially along one side of the sonotrode shaft. For all three tip positions, delamination only occurs for flakes close to the ``active regions'', as revealed by the optical imaging \textit{Results}.

Although not visible in figure~\ref{fig:camera_image_4cm2}~(d \& e), owing to the housing components shown in figure~\ref{fig:platform}~(b \& c), the anode flakes have also settled to the bottom of the tube transducer, prior to sonication. During the first $\qty{400}{\milli \second}$ of sonication, at both $\qty{55}{\watt}$ and $\qty{106}{\watt}$ input powers, the flakes remain on the bottom of the tube but graphite coating has clearly been delaminated. The liberated powder appears to gather along what may be the radial bubble filaments, identified by the HSI and SCL observations, in the lower half of the transducer bore. From $\qty{800}{\milli \second}$, delaminated copper foil flakes are lifted from the bottom and transported through the tube bore, and liberated powder concentrates at the central on-axis position. Figure~\ref{fig:camera_image_4cm2}~(e) indicates that the increase in tube transducer input power to $\qty{106}{\watt}$ increases the rate of flake delamination. Enhanced streaming also disperses liberated powder more evenly through the bore of the transducer, with a notable concentration at the central on-axis position, toward the end of the sonication. 

\begin{table*}[ht]
\centering
\caption{The mass removed from batches of LiB anode flakes with effective areas of $\qty{4}{\centi \metre \squared}$ and $\qty{17}{\centi \metre \squared}$, following $\qty{2}{\second}$ sonications in each of the transducer configurations. Data is provided as the mean of five separate flake batches $\pm$ standard deviation, for each configuration.}
\begin{tblr}{
  cells = {c},
  cell{1}{1} = {r=3}{},
  cell{1}{2} = {c=2}{},
  cell{2}{2} = {c=2}{},
  hline{1,9} = {-}{0.08em},
  hline{4} = {-}{},
}
\textit{Transducer configuration }   & \textit{Mass removed} ($\unit{\milli \gram}$) & \\
                                     & \textit{Effective area of LiB anode flakes} & \\
                                     & $\qty{4}{\centi \metre \squared}$  & $\qty{17}{\centi \metre \squared}$  \\
Sonotrode tip in top position                       & $ 8  \pm 4$   & $ 5   \pm 2$  \\
Sonotrode tip in centre position                    & $ 17 \pm 1$   & $ 50  \pm 15$ \\
Sonotrode tip in bottom position                    & $ 34 \pm 1$   & $ 170 \pm 26$ \\
Tube transducer at $\qty{55}{\watt}$ input power    & $ 89 \pm 3$   & $ 169 \pm 13$ \\
Tube transducer at $\qty{106}{\watt}$ input power   & $ 89 \pm 2$   & $ 386 \pm 2$ 
\end{tblr}
\label{tab:mass_removed}
\end{table*}

The degree of graphite coating delamination from flakes with a $\qty{4}{\centi\metre\squared}$ effective area, in each of the transducer configurations, is quantified in figure~\ref{fig:lib_compare_sonotrode_and_tube_4cm2} and table~\ref{tab:mass_removed}. The ``remaining mass percentage'' of figure~\ref{fig:lib_compare_sonotrode_and_tube_4cm2} is the mass of LiB anode flakes recovered from the liquid volume, following sonication, as a percentage of the pre-sonication mass. Each data point represents the average of five measurements, with error bars indicating the standard deviation. Complete removal of graphite coating from a batch of flakes results in a remaining mass value of $\qty{23}{\percent}$, indicated by the dashed line across the bar chart. A representative image of a batch collected from each transducer configuration is also presented in figure~\ref{fig:lib_compare_sonotrode_and_tube_4cm2}.

For top and centre sonotrode tip positions, remaining mass percentages of $\qty{94}{\percent}$ and $\qty{89}{\percent}$, confirm the minimal delamination observed in figure~\ref{fig:camera_image_4cm2}~(a \& b). Percentage remaining mass decreased to $\qty{72}{\percent}$ for the tip in the bottom position, in-line with the observation of figure~\ref{fig:camera_image_4cm2}~(c). The tube transducer at both input powers effectively removes all graphite coating from all batches of flakes. Inspection of the recovered flakes of figure~\ref{fig:lib_compare_sonotrode_and_tube_4cm2} reveals further differences in the effects of sonications by the sonotrode and tube transducer configurations. The morphology of flakes recovered following sonotrode treatment were largely unaffected and planar, while those recovered from the tube transducer were contorted and fragmented by the sonication.

\subsection{Effect of increasing anode flake loading on the degree of graphite coating delamination} \label{sec:results_effect_area}
\begin{figure*}[t]
\centering
\includegraphics[]{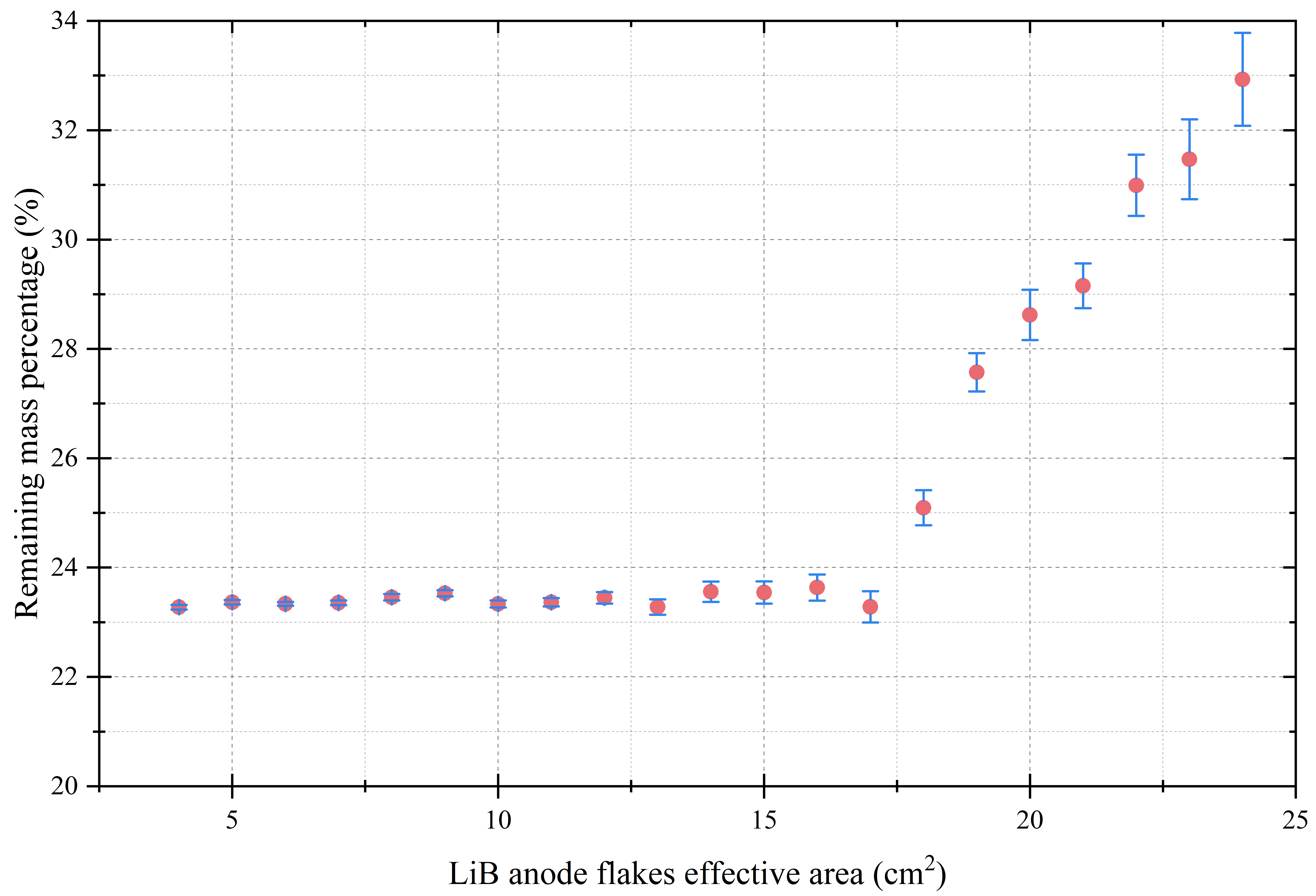}
\caption{The effect of LiB anode flake effective area on the remaining mass percentage recovered from the tube transducer, following $\qty{2}{\second}$ of sonication at $\qty{106}{\watt}$. A lower remaining mass percentage value indicates a higher degree of graphite coating delamination, with $\qty{23}{\percent}$ representing complete removal.}
\label{fig:tube_106W_flakes}
\end{figure*}

The effect of increasing the flake loading added to the tube transducer for $\qty{2}{\second}$ sonications at $\qty{106}{\watt}$ input power, was investigated with effective area increments of $\qty{1}{\centi \metre \squared}$, figure~\ref{fig:tube_106W_flakes}. Each data point represents the average measurements from five sonications, with error bars indicating the standard deviation. 

For effective area loading up to and including $\qty{17}{\centi\metre\squared}$ remaining mass percentages were between $\qty{23}{\percent}$ and $\qty{24}{\percent}$, corresponding to complete graphite coating removal. Remaining mass percentage thereafter increased with effective area load, to approximately $\qty{33}{\percent}$ at $\qty{24}{\centi\metre\squared}$. 

\subsection{Graphite coating removal from anode flakes of \texorpdfstring{$\qty{17}{\centi \metre \squared}$}{17-cm-squared} effective area } \label{sec:results_lib_17cm2}
\begin{figure*}[t]
\centering
\includegraphics[]{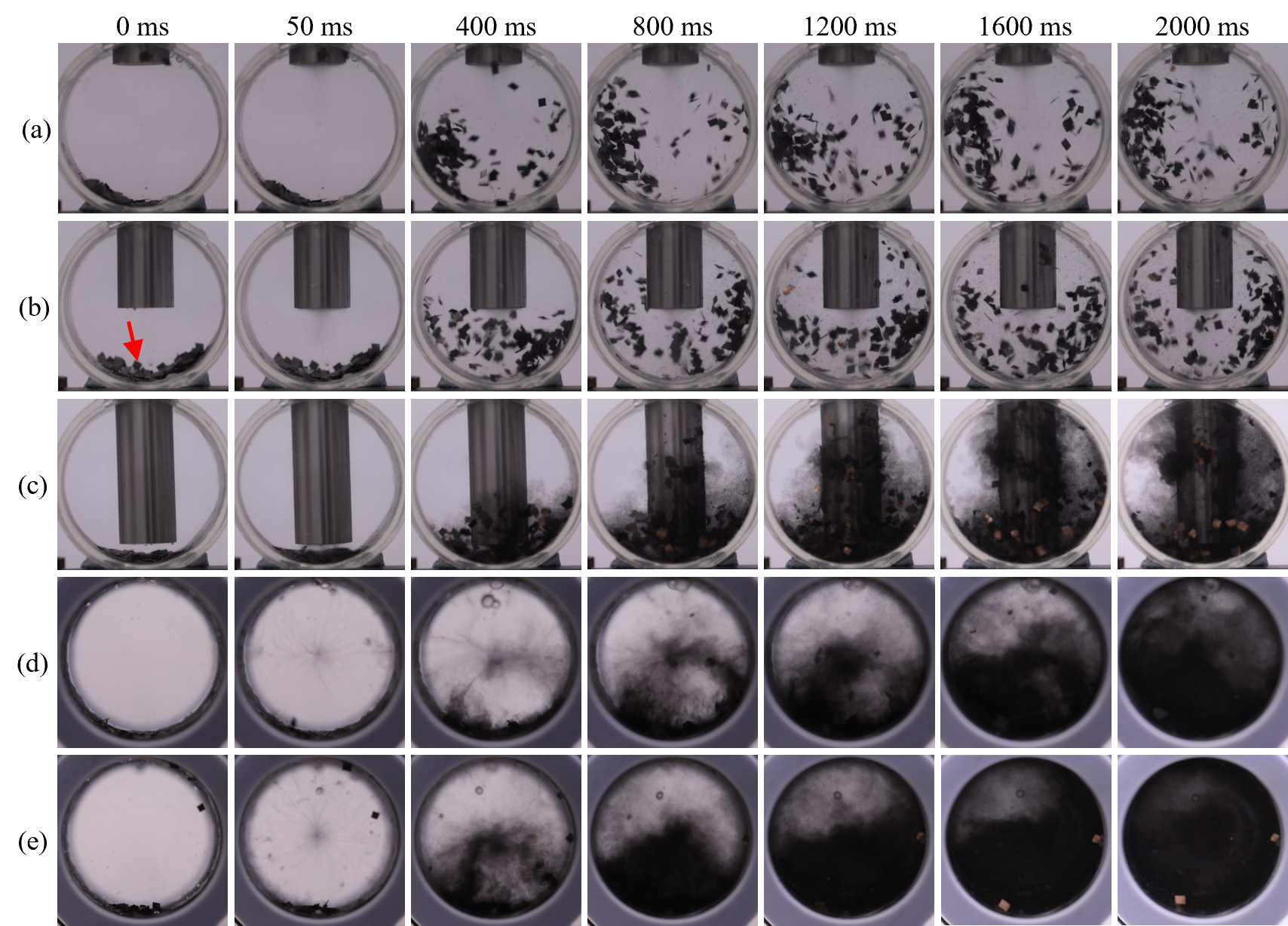}
\caption{Images from a conventional recording of graphite coating delamination from $\qty{17}{\centi\metre\squared}$ of LiB anode flakes, during a $\qty{2}{\second}$ sonication with the sonotrode tip in the (a)~top, (b)~centre and (c)~bottom position, and the tube transducer at (d)~$\qty{55}{\watt}$ and (e)~$\qty{106}{\watt}$ input power. Scale is provided by the $\qty{20}{\milli\metre}$ diameter sonotrode tip. The red arrow in (b) indicates the anode flakes settled at the bottom of the cylindrical vessel, prior to sonication.}
\label{fig:camera_image_17cm2}
\end{figure*}

\begin{figure*}[t]
\centering
\includegraphics[]{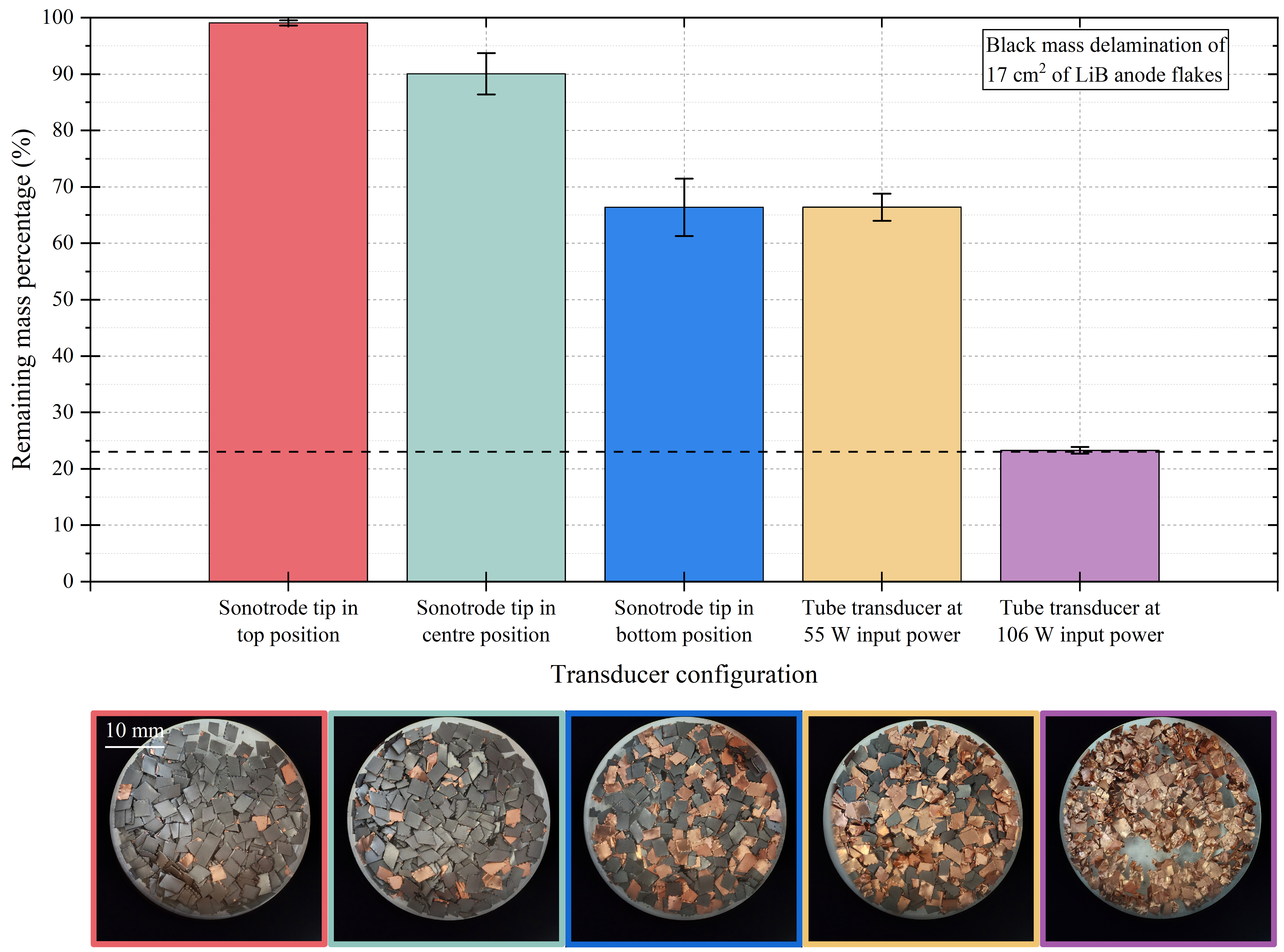}
\caption{Remaining mass percentage of LiB anode flakes with a $\qty{17}{\centi\metre\squared}$ effective area following $\qty{2}{\second}$ sonications in each of the five configurations. Complete graphite coating removal is indicated by a dashed line at $\qty{23}{\percent}$. Also shown are pictures of the  flake batches, post-sonication, for each configuration.}
\label{fig:lib_compare_sonotrode_and_tube_17cm2}
\end{figure*}

Figure~\ref{fig:tube_106W_flakes} reveals a threshold effective area of $\qty{17}{\centi \metre \squared}$ for complete removal of graphite coating from a batch of anode flakes. Observations of the removal process for batches of flakes of this threshold effective area, in the five transducer configurations, are reported in this section.

Figure~\ref{fig:camera_image_17cm2}~(a \& b) show that during a $\qty{2}{\second}$ sonication with the sonotrode at $\qty{55}{\watt}$ input power, with the tip in the top and centre positions, the flakes are agitated by the acoustic streaming induced by the sonication, similar to the lower effective area loading observations of figure~\ref{fig:camera_image_4cm2}. Once again, however, there is only slight darkening of the water indicating a low degree of delamination. With the tip in the bottom position, figure~\ref{fig:camera_image_17cm2}~(c), significant delamination occurs throughout the sonication. Clearly, the higher degree of darkening is due to more graphite coating powder being liberated from the higher number of flakes. The powder again tends to gather along the shaft of the sonotrode probe and delaminated copper flakes are apparent in the lower regions of the cylindrical vessel, later in the sonication.

Figure~\ref{fig:camera_image_17cm2}~(d \& e) depict the graphite coating removal process in the tube transducer at input powers of $\qty{55}{\watt}$ and $\qty{106}{\watt}$, respectively. Powder liberated from the flakes, initially at the bottom of the tube, is once more transported upwards through the bore, initially collecting at the central on-axis region. At both powers, the water is almost completely blackened by the end of the $\qty{2}{\second}$ sonication.

The graphite coating removed from $\qty{17}{\centi \metre \squared}$ flake batches in each of the transducer configurations are quantified in figure~\ref{fig:lib_compare_sonotrode_and_tube_17cm2} and  table~\ref{tab:mass_removed}. With the sonotrode tip in the top and centre positions $\qty{99}{\percent}$ and $\qty{91}{\percent}$ of the flake mass was retained. The sonotrode tip in the bottom position and the tube transducer with $\qty{55}{\watt}$ input power exhibited a similar degree of delamination, with both producing remaining mass percentages of $\qty{67}{\percent}$. The tube transducer at $\qty{106}{\watt}$ achieved complete removal, with a remaining mass percentage of $\qty{24}{\percent}$. 

Images of the recovered flakes in figure~\ref{fig:lib_compare_sonotrode_and_tube_17cm2} gives further insights to the delamination caused by each configuration, across the population of flakes. With the sonotrode tip in the top and centre positions, it can be seen that the little delamination that did occur came from complete removal of graphite coating from a small number of flakes. This supports the assertion that delamination only occurs when streaming transports flakes to the active cavitation zone immediately below the tip. The recovered flake images for the bottom tip position and tube transducer at $\qty{55}{\watt}$ indicate that although the percentage remaining mass is similar, the tube achieved more uniform delamination across the population of flakes. As for the lower effective area batches, the recovered flakes from the tube transducer show signs of significant cavitation damage, especially at $\qty{106}{\watt}$.
\FloatBarrier

\section{Discussion} \label{sec:discussion}
In this work, we detail the construction of a novel power ultrasonic transducer from a radially poled piezoceramic tube and compare the cavitation generated to that induced by a conventional sonotrode device, at similar frequency and input power. The HSI and SCL results indicate that the most active cavitation generated during a tube transducer sonication is found along the central axis, as anticipated for an inwardly focused, radial mode excitation. Cavitation is also generated throughout the bore of the tube, including along the emitting, inner curved wall, from which cavitating bubbles migrate inwards, particularly along radial filamentous bubble structures that form and feed the central cluster.

These observations suggest that the tube transducer cavitates liquids in a way that combines the advantages of the two common Langevin transducer deployments, described in the \textit{Introduction}. Cavitation at sonotrode-like intensity is distributed throughout the liquid bulk, similar to an ultrasonic cleaning bath. Moreover, the most active cavitation is removed from the emitting surface, in contrast to the cavitation at the tip of a sonotrode. This difference could be beneficial to applications such as food processing, where contamination from tip erosion is a significant concern that has limited development and industrial adoption \cite{Dio09, freitasContinuousContactContaminationfree2006}.

In all studied transducer configurations, the HSI and SCL results translated well to the LiB anode delamination case-study application, in terms of predicting the sonoprocessing capabilities of each. For the sonotrode delamination of an intact anode section, the constrained cavitating region at the sonotrode tip limits graphite coating liberation to an approximately $\qty{4}{\centi \metre \squared}$ circular area of the section. Similarly, the effectiveness of sonotrode delamination reduces rapidly as the vertical tip-to-anode distance is increased. We also note that two sided delamination can be achieved if the section is suspended within the liquid. This is likely advantageous compared to previous reports, where a sheet was affixed to a solid boundary during sonication, for which delamination from only the side closest to the tip was observed \cite{lei2021lithium, lei2024effect}. For the same electrical input power of $\qty{55}{\watt}$, the tube transducer removes effectively all graphite coating from both sides of the $\qtyproduct{3 x 3}{\centi \metre}$ section, when it is suspended within the bore. This result further confirms that active cavitation is generated throughout the tube transducer with an intensity that can be harnessed for sonoprocessing applications. Indeed, the cavitation intensity generated by the tube at both input powers and, to a lesser extent, the sonotrode tip in close proximity to the anode samples, is likely too high for this particular case study application. The pitting and puncturing of the copper foil for these configurations is not ideal for reuse and recycling purposes. Specific applications will benefit from sonication parameter optimisation, including input voltage and pulsing (burst duration and duty cycle), to avoid overtreatment and minimise power consumption.

A primary motivation for developing the tube transducer is the relative ease with which it may be incorporated within a flow configuration. We envisage a modular configuration whereby multiple tube transducers, embedded within a pipe, subject flowing liquid to sonication as it passes through each tube transducer. For such a system, an application such as LiB electrode delamination would require the use of anode flakes, rather than large intact sheets. Shredded or crushed material would be suspended in a liquid or solvent of choice for flowing through a pipe that incorporates as many tube transducers as required by the demands of the particular application. 

In this work, however, all treated liquids were initially static, with all observed liquid motion due to acoustic streaming induced by the sonication. One consequence of this is that at the time of sonication initiation, flakes had settled to the bottom of the cylindrical vessel or tube transducer, under the action of gravity. For either effective area of flakes reported above, sonotrode sonications with the tip located in the top position produced negligible delamination, with only slightly more observed for when the tip was in the central position. Any delamination observed for these tip locations was due to streaming effects transporting flakes to, and through, the constrained cavitating region below the tip. 

Sonotrode sonications with the tip in the bottom position meant that the constrained cavitating region at the tip was in close proximity to the flakes from the outset of the sonication. For the lower effective area of flakes, streaming effects acted to remove flakes from this region, resulting in less delamination than might be expected. For the higher effective area, however, the extra mass of flakes at the bottom appears to have suppressed the streaming induced motion, such that a greater proportion of flakes were subjected to the cavitation within the constrained region. Indeed, this sonotrode configuration produced a similar degree of delamination as the tube transducer at the same electric input power. For the tube transducer, however, the initial location of the flakes at the bottom of the tube is not ideal for exploiting the on-axis region of highest intensity cavitation activity. Nonetheless, the tube transducer at the higher electrical input power of $\qty{106}{\watt}$ provided effectively complete delamination of both intact anode sections and flake samples with effective areas up to $\qty{17}{\centi \metre \squared}$---more than four times greater than the area treated by the sonotrode, on an intact section---over a $\qty{2}{\second}$ sonication.

In a flow-pipe configuration, flakes (or other particles of shredded material) may be expected to pass through the embedded tube transducers, suspended within the flowing liquid. The characterisation of the cavitation field generated would suggest that this will further enhance the performance of the tube transducer, over the static liquid results presented here, for which we relied on streaming effects to transport flakes off the bottom. In flow, flakes will enter the tube transducer distributed across the cross-sectional area of the bore, and will thereby be subjected to the distributed cavitation field, including the most intense on-axis region. Such a system will clearly require substantial optimisation, in terms of system parameters including (but not limited to) flow rate, shred size and concentration, number and spacing of tube transducers along the pipe, as well as sonication parameters including frequency, input power, and pulsing protocols for the transducers. 

\section{Conclusion} \label{Conclusion}
An inwardly focusing tube transducer is described as a new source device for power ultrasonics. We show that the tube transducer generates cavitation activity at sonotrode-like intensities but distributed through the bore of the tube, with peak activity at the central axial position. We demonstrate that sonication with the tube transducer offers several advantages over sonotrode sonications for a case study application of graphite coating delamination from lithium-ion battery anode. Ongoing and future work will develop this technology for high throughput flow-based sonoprocessing and sonochemistry applications.

\section*{CRediT authorship contribution statement}
\textbf{Shida Li:} Writing – review \& editing, Writing – original draft, Investigation.
\textbf{Paul Daly:} Writing – review \& editing, Writing – original draft.
\textbf{Ben Jacobson:} Writing – review \& editing, Investigation.
\textbf{Joshua Cooke:} Writing – review \& editing, Investigation.
\textbf{Chunhong Lei:} Writing – review \& editing, Investigation.
\textbf{Andrew P. Abbott:} Writing – review \& editing, Supervision, Project administration, Funding acquisition.
\textbf{Andrew Feeney:} Writing – review \& editing, Supervision, Project administration, Funding acquisition.
\textbf{Paul Prentice:} Writing – review \& editing, Supervision, Project administration, Funding acquisition, Conceptualisation.

\section*{Declaration of Competing Interest}
The Glasgow-based authors declare that the technology described in this paper is the subject of a pending UK patent application, filed by the University of Glasgow.

\section*{Acknowledgements}
This work received financial support from the UK Engineering and Physical Sciences Research Council (EPSRC) through the SonoCat project (grant EP/W018632/1), the EU-funded APOLLO project (grant agreement ID 101122277 and Innovate UK guarantee no. 10125993), the UKRI-funded REACT program, and the Faraday Institution (award numbers FIRG027, FIRG057 and FIRG085).

\appendix
\section{Supplementary data}
\label{app1}
Supplementary data to this article can be found online.

\bibliographystyle{ieeetr}
\bibliography{ref}
\end{document}